\documentclass{article}
\usepackage{times}
\usepackage{epsfig,amssymb,amsfonts,amsmath}

\def\eeq{\end{equation}}
\def\beq{\begin{equation}}
\def\bea{\begin{eqnarray}}
\def\eea{\end{eqnarray}}
\begin{document}

\title{Analysis of return distributions in the coherent noise model}
\author{Ahmet Celikoglu$^1$, 
Ugur Tirnakli$^{1,2,}$\thanks{ugur.tirnakli@ege.edu.tr} $\,$
 and S\'{\i}lvio M. Duarte Queir\'{o}s$^{3}$ \\ \\
$^1$ Department of Physics, Faculty of Science, \\
Ege University, 35100 Izmir, Turkey \\
$^2$ Division of Statistical Mechanics and Complexity, \\
Institute of Theoretical and Applied Physics (ITAP),\\
Kaygiseki Mevkii, 48740 Turunc, Mugla, Turkey \\
$^3$ Centro de F\'{\i}sica do Porto, Rua do Campo Alegre 687,\\
4169-007 Porto, Portugal}

\date{\today}

\maketitle

\begin{abstract}
The return distributions of the coherent noise model are studied for the system
size independent case. It is shown that, in this case, these distributions
are in the shape of $q$-Gaussians, which are the standard distributions
obtained in nonextensive statistical mechanics. Moreover, an exact relation
connecting the exponent $\tau$ of avalanche size distribution and the $q$
value of appropriate $q$-Gaussian has been obtained as $q=(\tau +2) / \tau$.
Making use of this relation one can easily determine the $q$ parameter values of
the appropriate $q$-Gaussians \emph{a priori} from one of the well-known
exponents of the system. Since the coherent noise model has the advantage of
producing different $\tau$ values by varying a model parameter $\sigma$,
clear numerical evidences on the validity of the proposed relation have been
achieved for different cases. Finally, the effect of the system size has also
been analyzed and an analytical expression has been proposed, which is
corroborated by the numerical results.
\end{abstract}

%\begin{keyword}
%\PACS  05.20.-y \sep 05.70.Ln \sep 89.70.Cf  
%\end{keyword}
%\end{frontmatter}

\newpage

%%%%%%%%%%%%%%%%%%%%%%%%%%%%%%%%%%%%%%%%

\section{\label{sec:Int}Introduction}

%%%%%%%%%%%%%%%%%%%%%%%%%%%%%%%%%%%%%%%%%
Throughout the last two decades the interest in extended dynamical systems
has experienced a steady increase. These systems exhibit avalanches of
activity whose size distributions are of power-law type. Although there is
not a unique nor unified theory which totally explains all the features of
these complex systems, there exist several known mechanisms producing
power-law behavior. One of the most popular and well-studied mechanisms is
that of self-organized criticality (SOC) introduced by Bak, Tang and
Wiesenfeld \cite{SOC}. Many physical systems and models have shown to
exhibit SOC \cite{books}. The most important feature of all these systems is
that the entire system is under the influence of a small local driving
force, which makes the system evolve towards a critical stationary state
having no characteristic spatiotemporal scale, without invoking a
fine-tuning of any parameter. On the other hand, SOC is not the only
mechanism causing power-law correlations that appear in a nonequilibrium
steady state. Another simple and robust mechanism exhibiting the same feature 
in the absence of criticality is the coherent noise model (CNM) \cite{newman1,CNM}. 
The CNM is based on the notion of an external stress acting coherently onto all 
agents of the system without having any direct interaction with agents. 
Therefore, the model does not exhibit criticality, but it still gives a 
power-law distribution of event sizes (avalanches).

Recently, it was presented an analysis method to interpret SOC behavior in
the limited number of earthquakes from the World and California catalogs by
making use of the return distributions (\textit{i.e.,} distributions of the
avalanche size differences at subsequent time steps) \cite{caruso}. In their
work Caruso \textit {et al} obtained the first evidence that the return distributions
seem to have the form of $q$-Gaussians, standard distributions appearing
naturally in the context of nonextensive statistical mechanics \cite%
{tsallis1,tsallis2}. Based on the assumption that there is no correlation
between the size of two events, they were also able to propose a relation
between the exponent $\tau $ of the avalanche size distribution and the $q$
value of the appropriate $q$-Gaussian as
\begin{equation}
q=e^{1.19\;\tau ^{-0.795}}\;\;,  \label{qvstau1}
\end{equation}%
which is rather important since it makes the $q$ parameter determined
\textit{a priori} and therefore it acquits $q$ of becoming a fitting parameter.
The only little drawback of their work was that the number of data taken from
the catalogs is not sufficiently large to obtain a very precise $\tau $
exponent and also clear return distributions with well-defined tails (which
is important in order to verify how good the distribution approaches a
$q$-Gaussian). Consequently, Eq.~(\ref{qvstau1}) could not be rigorously tested
until a very recent effort by Bakar and Tirnakli in \cite{bbtt}, where the same
analysis was made using a simple SOC model known as the Ehrenfest dog-flea model
in the literature \cite{ehrenfest} (see also \cite{nagler1,nagler2}). Thanks
to the simplicity of the dog-flea model, it was possible to achieve
extensive simulations with very large system sizes (up to $10^{7}$) and also
very large number of data elements (up to $2\times 10^{9}$).
Accordingly, from these extensive simulations, it was obtained a value of
$\tau =1.517$, which is in accordance with the ``mean-field'' exponent $3/2$
determined in several problems~\cite{kac,reviewtime,celia}.
Thence the $q$ value of return distributions was deduced {\it a priori}
from Eq.~(\ref{qvstau1}).

In this work, we plod along this way by setting forth the following points:
(i)~first, we will obtain an exact relation between $\tau $ exponent of the
avalanche size distribution and the $q$ value of the appropriate
$q$-Gaussian without resorting to any assumption and compare it to
Caruso \textit {et al} relation given in Eq.~(\ref{qvstau1}),
(ii)~since the CNM has the advantage of producing different $\tau $ values
by varying a model parameter $\sigma $ \footnote{In the dog-flea model there
is only one available value of $\tau$ since the only parameter is the
number of fleas.}, we now have the opportunity to
test the validity of our exact relation (and also the Caruso \textit {et al}
relation) not only for one case but for various cases, (iii)~since the
corresponding return distributions are expected to converge to the $q$-Gaussian
as the system size goes to infinity, the effect of finite system size is
also important and we shall try to analyze this effect proposing an
analytical expression, (iv)~and finally since this model is not a SOC model,
our results also give us the possibility of checking the generality of this
behavior observed so far in SOC models.

%%%%%%%%%%%%%%%%%%%%%%%%%%%%%%%%%%%%%

\section{\label{sec:Model} The coherent noise model}

%%%%%%%%%%%%%%%%%%%%%%%%%%%%%%%%%%%%%
Let us start by introducing the CNM. It is a system of $N$ agents, each one
having a threshold $x_{i}$ against an external stress $\eta $. The threshold
levels and the external stress are randomly chosen from probability
distributions $p_{thresh}(x)$ and $p_{stress}(\eta )$, respectively.
Throughout our simulations we use the exponential distribution for the
external stress, namely, $p_{stress}(\eta )=(1/\sigma )\exp (-\eta /\sigma )$
and the uniform distribution ($0\leq x\leq 1$) for $p_{thresh}(x)$. The dynamics
of the model is very simple: (i)~generate a random stress $\eta $ from $%
p_{stress}(\eta )$ and replace all agents with $x_{i}\leq \eta $ by new
agents with new threshold drawn from $p_{thresh}(x)$, (ii)~choose a small
fraction $f$ of $N$ agents and assign them new thresholds drawn again from $%
p_{thresh}(x)$, (iii)~repeat the first step for the next time step. The
model can be described in the form of a two step-master equation that we present
in the appendix. The number of agents replaced in the first step of the
dynamics determines the event size $s$ for this model. Although the CNM has
been introduced for analyzing biological extinctions \cite{newman1}, it has
then been adopted as a very simple mean field model for earthquakes even
though no geometric configuration space is introduced in the model \cite{CNM}.
It is shown that the model obeys the Omori law for the temporal decay
pattern of aftershocks \cite{wilke}, exhibits aging phenomena \cite{CNMaging}
and power-law sensitivity to initial conditions \cite{ebru}.

%%%%%%%%%%%%%%%%%%%%%%%%%%%%%%%%%

\section{\label{sec:PDF} Avalanche Size and Return Distributions}

%%%%%%%%%%%%%%%%%%%%%%%%%%%%%%%%%

\subsection{Size independent case}

As pointed out in \cite{CNM}, there is advantage in choosing the uniform distribution
($0\le x\le 1$) for the thresholds of the CNM agents seeing that the model can be
simulated in the $N\rightarrow\infty$ limit using a fast algorithm which acts
directly on the threshold distribution instead of acting on the agents of the
system. This enables us to obtain the avalanche size distribution $P(s)$ of the
model as being independent of the system size. The distribution $P(s)$ is expected
to be a power-law over many decades until the $s$ values reach a particular point
$s\sim\sigma$, thereafter it falls off exponentially. From our point of view this is
rather important since it means that if we measure the avalanche size exponent
$\tau$ using the region $s<\sigma$, then we must use this $\tau$ value to predict
\textit{a priori} the $q$ value of the $q$-Gaussian that the return distribution is
expected to converge in the \textit{entire region} without any deterioration
(not only in the central part but also in the tails). The results obtained for the
avalanche size distributions of three representative cases with $\sigma=0.01$,
$\sigma=0.05$ and $\sigma=0.065$ are given in the left column of Fig.~1.

%------------------------------ Figure 1 ----------------------------
\begin{figure*}[t]
\begin{center}
\includegraphics[width=6cm]{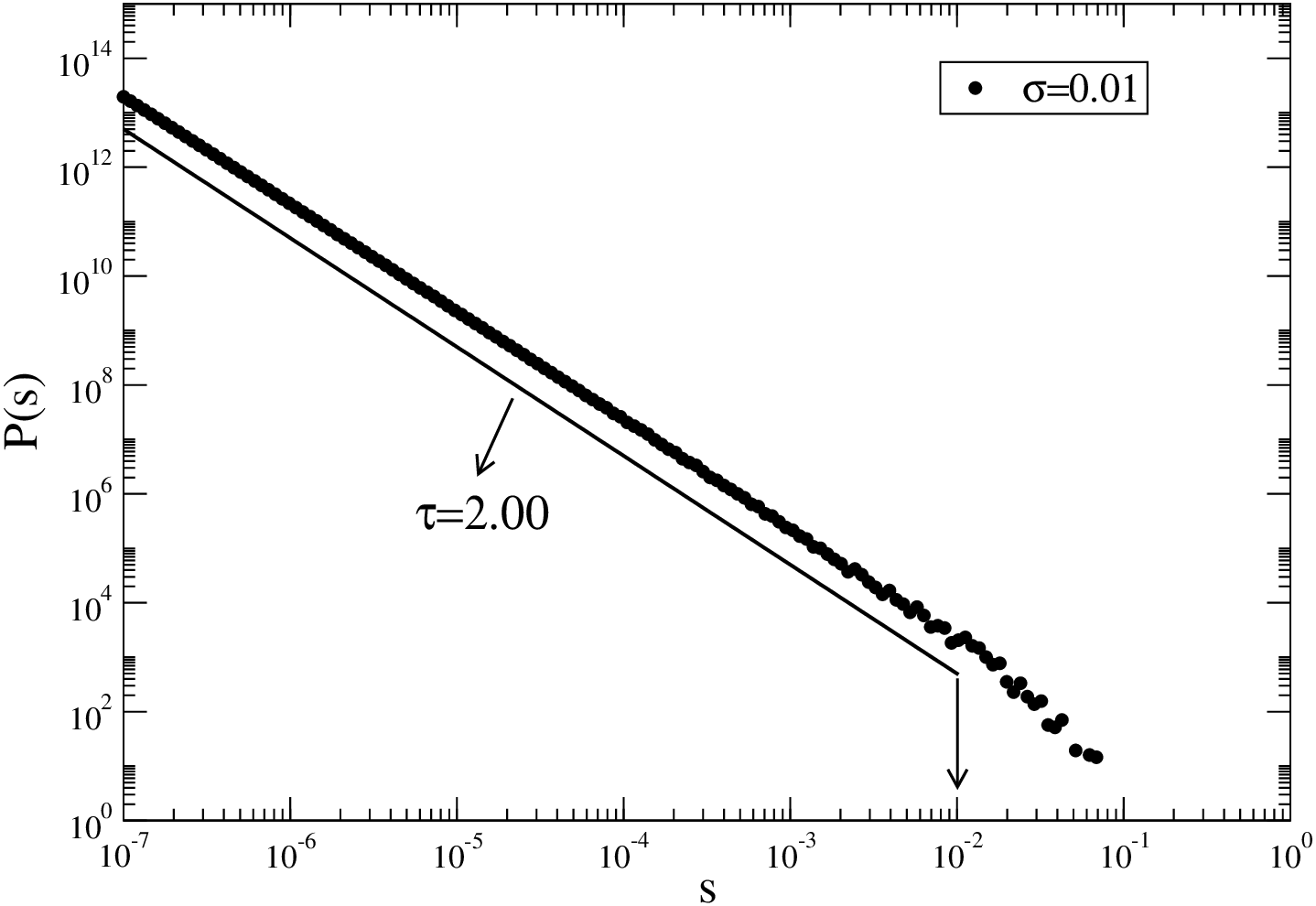} %
\includegraphics[width=6cm]{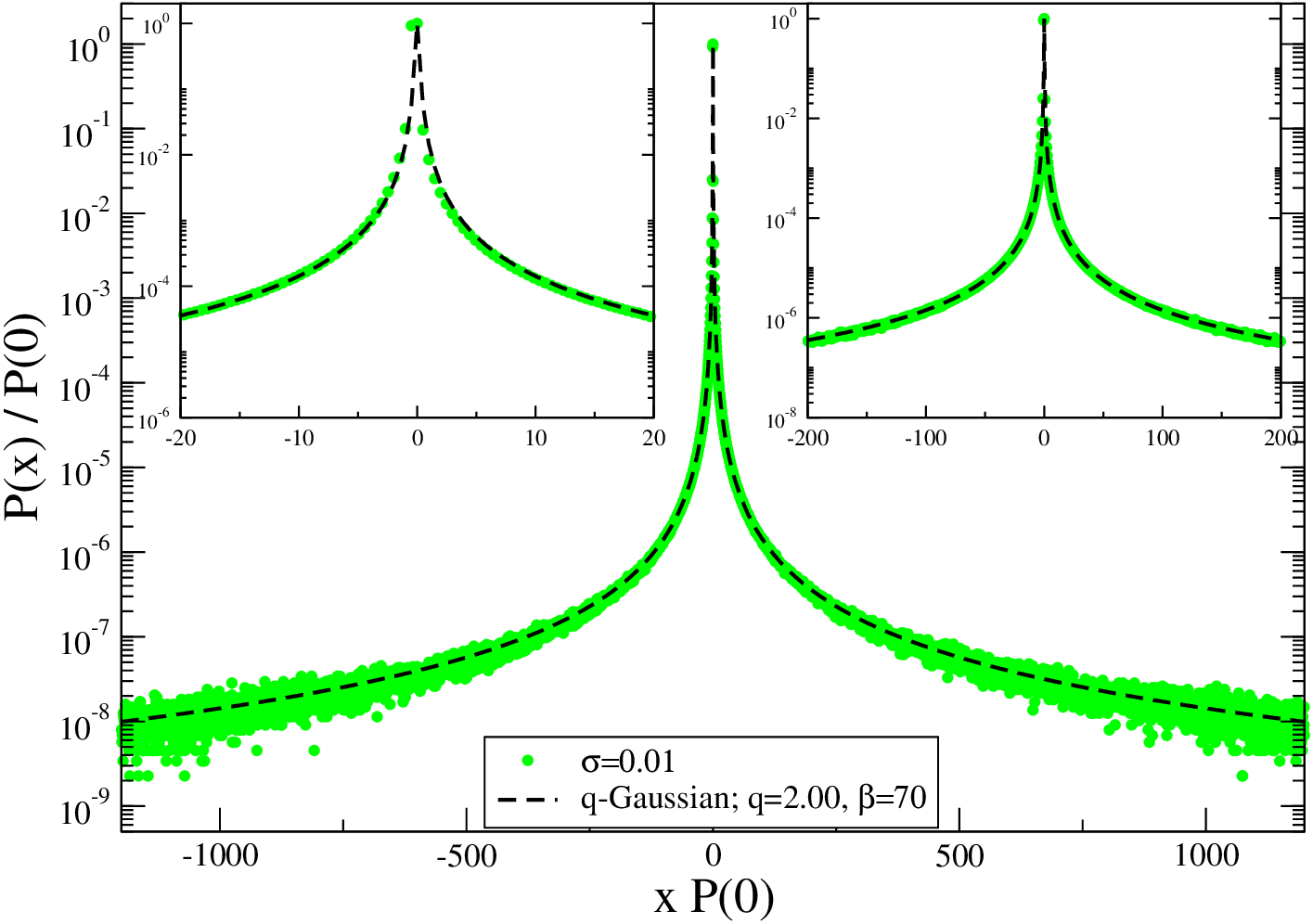} \\[1cm]
\includegraphics[width=6cm]{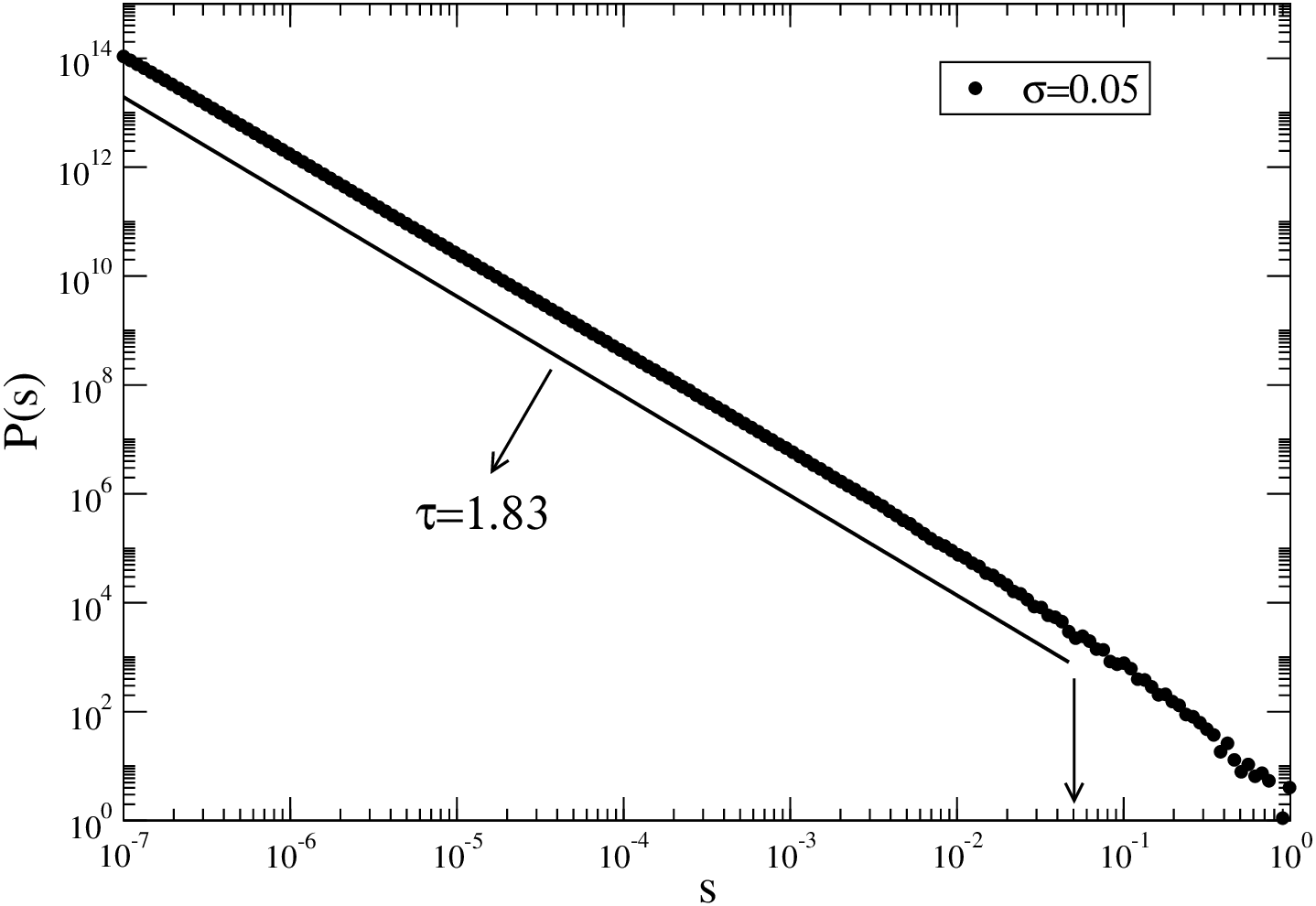} %
\includegraphics[width=6cm]{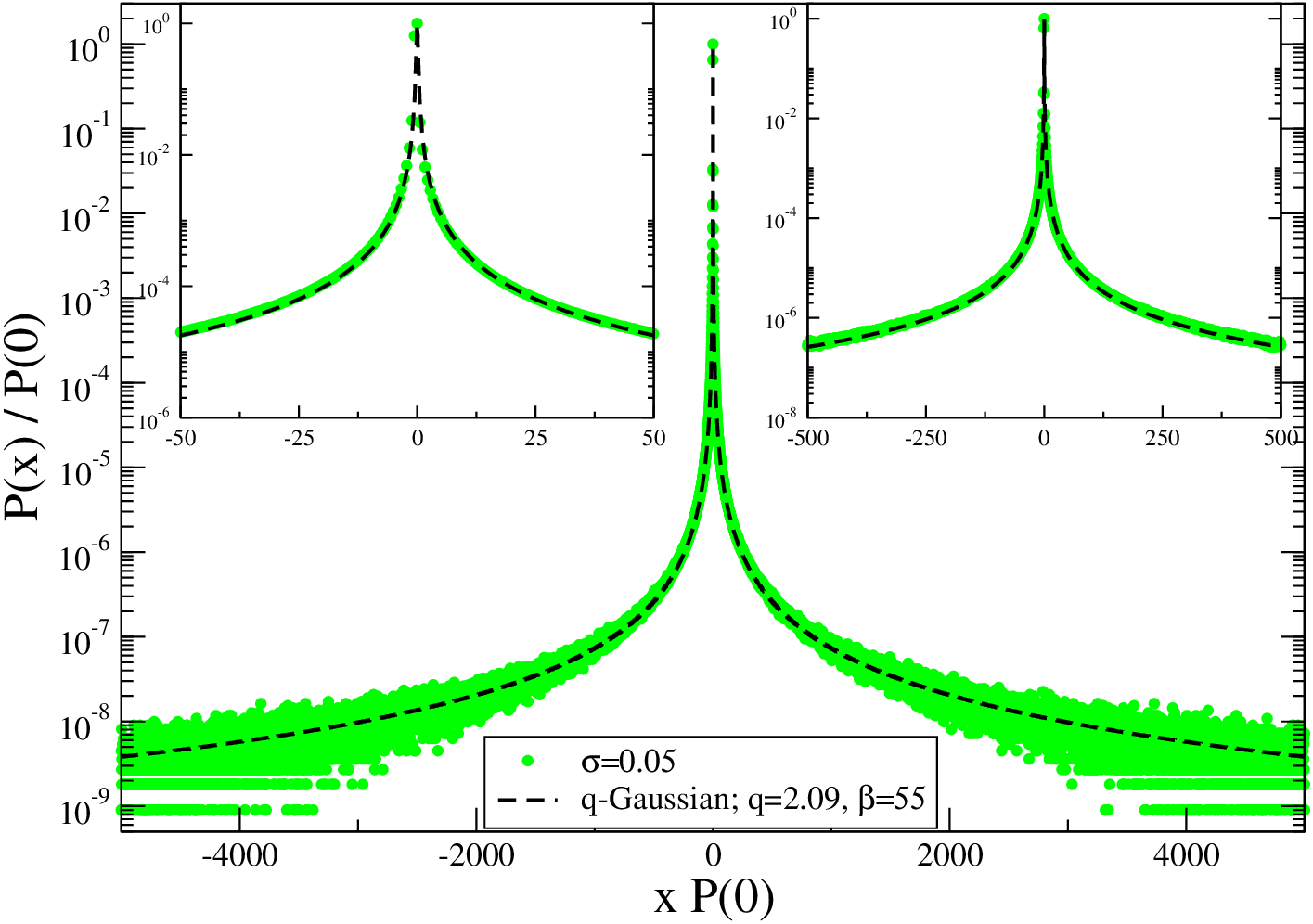} \\[1cm]
\includegraphics[width=6cm]{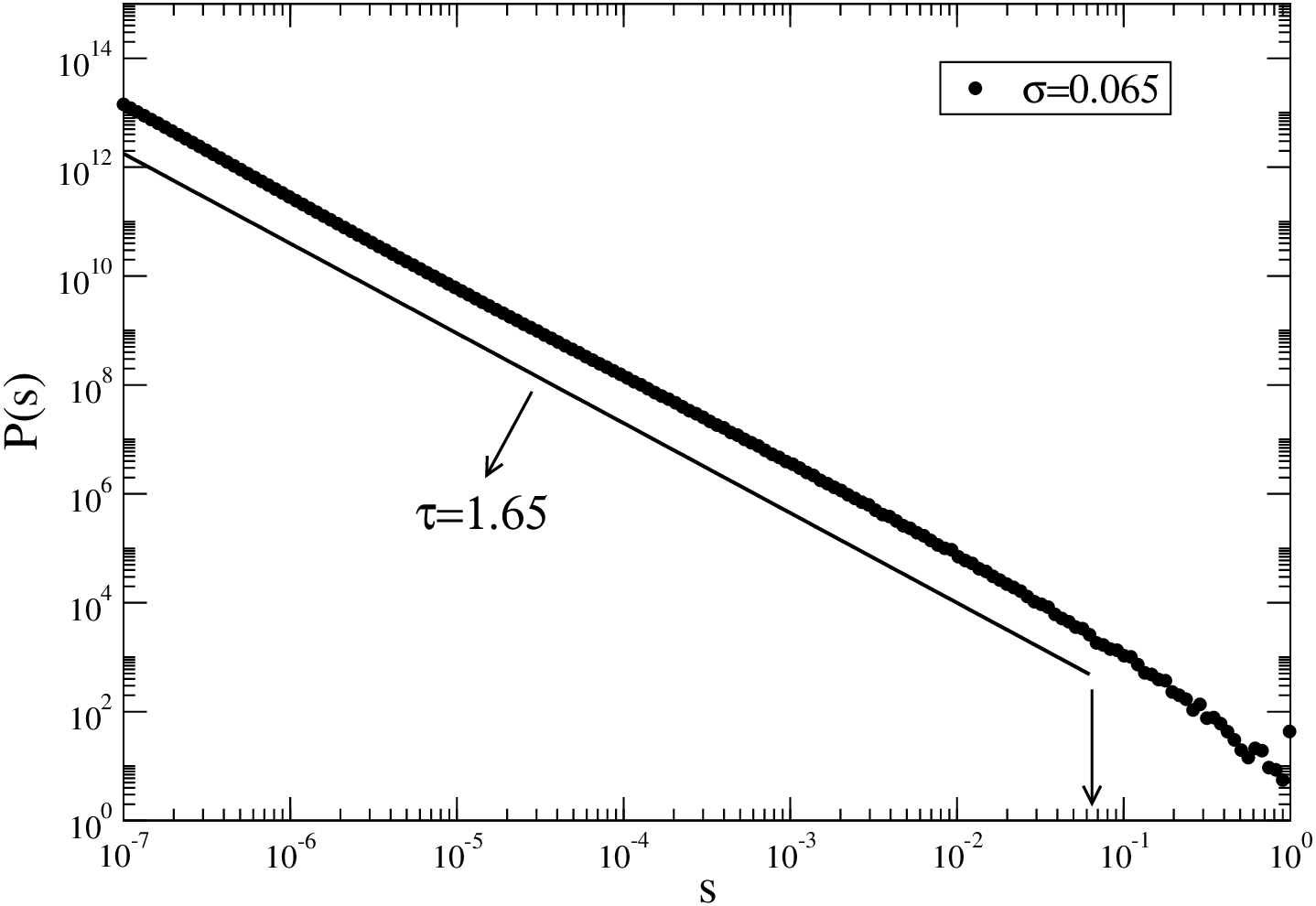} %
\includegraphics[width=6cm]{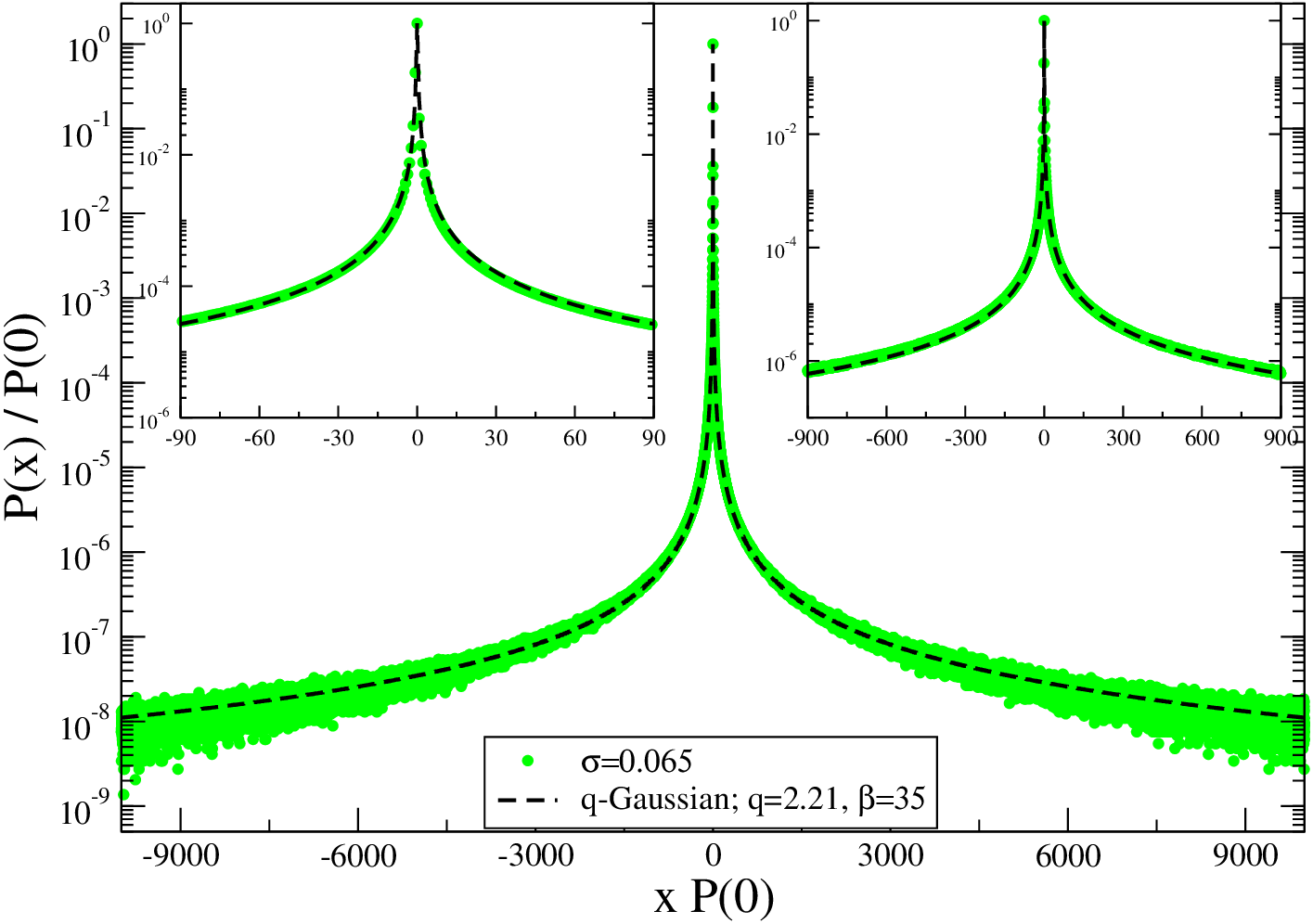}
\end{center}
\caption{Left column: Avalanche size distributions of three
representative values of $\protect\sigma$. For each case, the $\protect\tau$
value is calculated using standard regression method for the region
$s<\protect\sigma$. Right column: Return distributions for the same three
cases. Two zooms of the central part are given in the insets for better
visualization. For each case, $f=10^{-7}$ and $2\times 10^9$ experiments are
generated.}
\label{fig:fig1}
\end{figure*}

Since each case with different $\sigma $ values has a different size
exponent $\tau $, this allows us to check the validity of Caruso \textit {et al}
relation given in Eq.~(\ref{qvstau1}) or any other equation relating $\tau $
values to the $q$ values of the appropriate $q$-Gaussians. From the
master-equation of the CNM is theoretically possible to compute the
probability of $s$ and bringing to bear standard techniques~\cite{reviewtime} to obtain the return
distribution. However, its level of complexity turns out the solution almost
analytically impossible or its (asymptotic) behavior deeply unclear as it happens
in several other problems of this class~\cite{bugflea}. Regardless, we are in the
position where we can propose an exact relation for the return distribution
$P\left( \Delta s\right) $ bringing into play no other assumption than the
distribution of avalanche sizes, where $\Delta s$ is the difference between two
consecutive event sizes, i.e., $\Delta s=s(t+1)-s(t)$.
Let us mathematically define the avalanche size distribution,

\begin{equation}
p(s)\propto \left( \varepsilon +s\right) ^{-\tau },\qquad
\left( \tau>1\right) ,
\label{ps}
\end{equation}
with $\varepsilon $ being a constant value describing the asymptotic limit
$s\rightarrow 0$.
The process of avalanches is completely Markovian (independent) and therefore
the probability of the difference of sizes $\Delta s$ is
\begin{eqnarray*}
P\left( \Delta s\right) &=&\int_{0}^{\infty } \int_{0}^{\infty }p(s)\,p(s^{\prime
})\,\delta \left( \Delta s -\left( s-s^{\prime }\right) \right)
\,ds^{\prime }\,ds. \\
&=&\int_{0}^{\infty }\left( \varepsilon +s\right) ^{-\tau }\left(
\varepsilon +\Delta s+s\right) ^{-\tau }\,\Theta \left( \Delta s+s\right)
\,ds,
\end{eqnarray*}
where $\Theta(\ldots)$ is the Heaviside step function and $s^{\prime }$ denotes
the previous avalanche size.
Making use of \cite{gradshteyn} and attending to the symmetric nature of
$P\left(\Delta s\right)$ we can explicit the negative branch,
\begin{equation}
P\left( \Delta s\right) =\left\vert \Delta s\right\vert ^{1-2\,\tau }\left( B%
\left[ \frac{\varepsilon }{\Delta s},1-\tau ,1-\tau \right] \left( -1\right)
^{\tau }+\mathcal{C}\left( \tau \right) \right) ,
\label{pdeltae1}
\end{equation}
where $\mathcal{C}\left( \tau \right) $ is a coefficient only depending on
$\tau$ and related to the convolution of very large values of $s$ with very
large values of $- s^{\prime }$ yielding a
$\left\vert \Delta s\right\vert^{1-2\,\tau }$
dependence \footnote{This can be flatly checked out performing the calculation
with $\varepsilon=0$.}.
Thus, the distribution is mainly described by the product of the isolated factor
by the incomplete Beta function $B[\ldots ]$.
Applying the asymptotic behavior $x \rightarrow 0$ of $B[x,a,b]$~\cite{functions}
we finally get,
\[
P\left( \Delta s\right) \sim \left\vert \Delta s\right\vert ^{-\,\tau
},\qquad \left( \Delta s\gg 1\right) .
\label{Pdeltas1}
\]
Taking note of the $q$-Gaussian distribution,
\begin{equation}
P(\Delta s)=P(0)\left[ 1+{\bar{\beta}}(q-1)\Delta s^{2}\right] ^{1/(1-q)},
\label{qgaussian}
\end{equation}
we straightforwardly obtain
\begin{equation}
q=\frac{\tau +2}{\tau } .
\label{qvstau2}
\end{equation}
This relation is slightly different from the approximate relation presented
in \cite{caruso} as can be seen in Fig.~2. For $\tau \rightarrow \infty$,
both relations approach $q = 1$ and they are almost identical except in the
region where $\tau $ values are smaller than $1.5$. Moreover, only the relation
(\ref{qvstau2}) correctly achieves $q=3$ value when $\tau =1$.
These values define the limits of the domain of each parameter so that the
distributions (\ref{ps}) and (\ref{qgaussian}) are normalizable.
The approximate relation Eq.~(\ref{qvstau1}) does not fulfill this condition
as $q\left( \tau =1\right) >3$. It should be noted that, since this discrepancy
is only meaningful for $\tau <1.5$, the approximate relation predicts the $q$ values
with a $| 0.01 |$ difference from the exact one, which are also acceptable for
all the cases we present.

%------------------------ Figure 2 --------------------------
\begin{figure*}[t]
\begin{center}
\includegraphics[width=9cm]{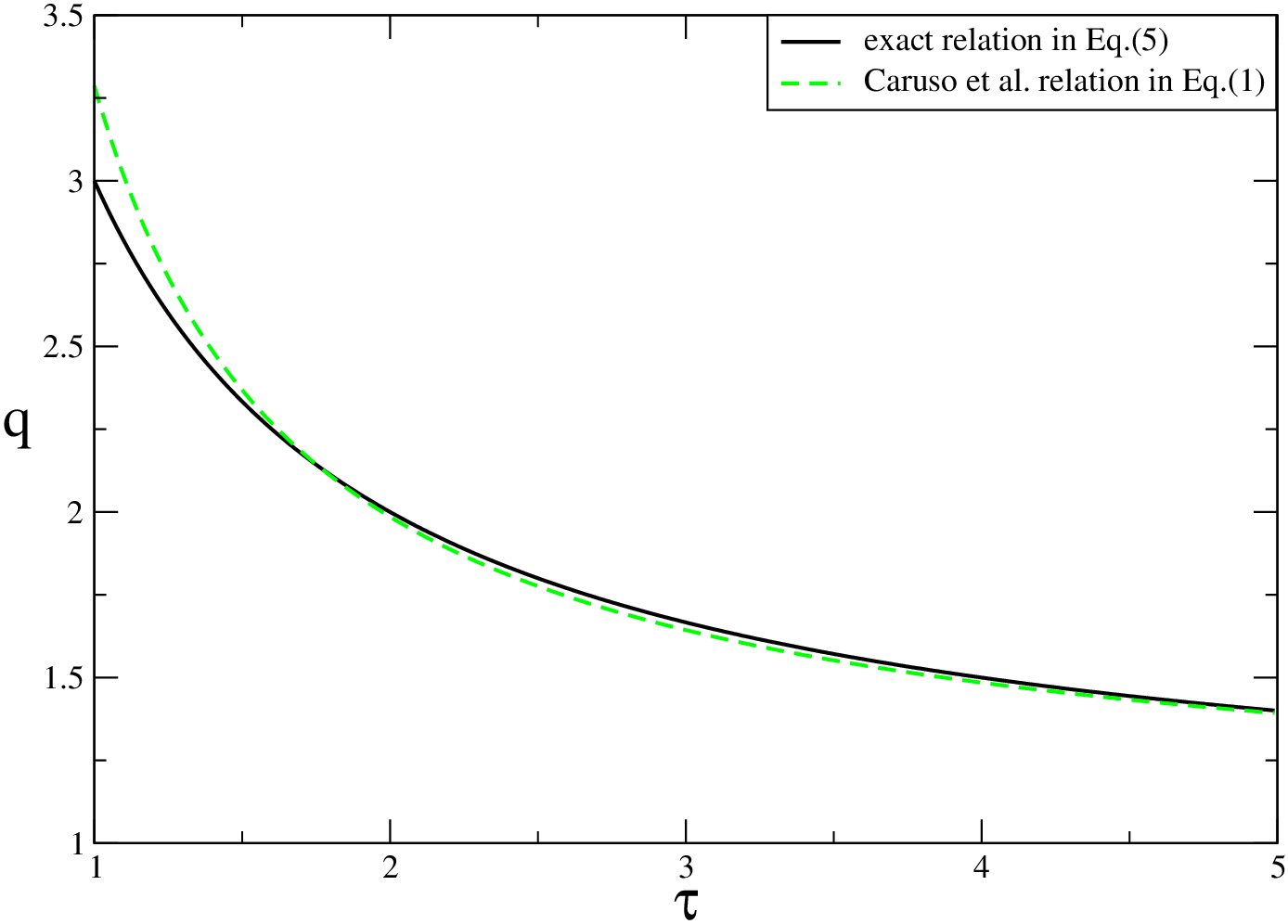}
\end{center}
\caption{Comparison of our exact relation (Eq.~(\protect\ref%
{qvstau2})) with that of Caruso \textit {et al} (Eq.~(\protect\ref{qvstau1})).}
\label{fig:fig2}
\end{figure*}
%-----------------------------------------------------------

We are now ready to proceed analyzing the return distributions.
The centered returns are given in terms of variable $x$

\begin{equation}
x=\Delta s-\left\langle \Delta s\right\rangle \;\;,
\end{equation}%
where $\left\langle ...\right\rangle $ represents the mean value of a given
data set. As can be seen from the right column of Fig.~1, in our simulations
we generated the return distributions of the three representative cases of
the CNM in order to check the validity of the relation (\ref{qvstau2}). In
each case, an extremely large number of events ($2\times 10^{9}$) has been
used to build the numerical distribution, namely, the central part and
tails. It is clear that the return distribution (green dots) can by no means
be approached by a Gaussian. They actually exhibit fat tails which agree
with $q$-Gaussians Eq. (\ref{qgaussian}) where ${\bar{\beta}}$ characterizes
the width of the distribution and $q$ is the parameter which should be
determined directly from Eq.~(\ref{qvstau2}) \textit{a priori} and therefore
is no longer a fitting parameter. In each panel on the right column of
Fig.~1, the dashed black lines represent the appropriate $q$-Gaussian with
the $q$ value obtained from Eq.~(\ref{qvstau2}). Perfect agreement with the
data can be easily appreciated not only for the tails but also for the
intermediate and the very central part as it is demonstrated in the insets.

%%%%%%%%%%%%%%%%%%%%%%%%%%%%%%%%%%%%%

\subsection{Size dependent case}

%%%%%%%%%%%%%%%%%%%%%%%%%%%%%%%%%%%%%
Although we might think that the size independent (i.e., infinite size) case
would be enough for such an analysis, we believe that it is still instructive
to look also at the size dependent case at least from two different
perspectives: (i)~we can check how the size of the system affects the shape
of the return distributions and whether the tendency is consistent with the
infinite size case as the size of the system increases, (ii)~unlike the CNM,
generic size independent cases cannot be achieved for such systems and thus
the only possibility is to always analyze the size dependent case.

It is very easy to implement the size dependent algorithm for the CNM. We
just need to apply the previously described steps of the dynamics to a system
of $N$ agents. As $N$ increases, this algorithm clearly slows down and for
the same number of events ($10^{9}$) the larger value of $N$ that we
can simulate in a reasonable time is $20000$. In Fig.~3(a) the behavior of
the avalanche size distribution is given for $\sigma =0.05$ case for various
$N$ values. It is clearly seen that the power-law regime is always followed
by an exponential decay of all the curves and this decay is postponed to larger
sizes as $N$ increases. For each $N$ case, we estimate the $\tau $ value
using the standard regression method in the region before the exponential
decay (we determine the size of this interval by looking at the regression
coefficient to become always more than 0.9997 in each case). Therefore, we
should expect that the exponential decay part would tamper with the
$q$-Gaussian behavior of the return distributions and this meddling must
diminish as $N$ gets larger and larger, which is in fact observed
in Fig.~3(b) for the return distributions of four representative $N$ values.
When $N$ values are very small, avalanche size distribution has a very short
power-law region and the exponential decay part dominates, which simply causes
the return distributions to deviate immediately from the $q$-Gaussian shape.
As $N$ increases, return distributions start approaching the thermodynamic
limit (dotted black line), which is a full $q$-Gaussian with $q=2.09$,
yielding better and better from the central part to the tails, i.e.,
as the expected scale-free regime sets in.

In order to explain this gradual approach to $q$-Gaussians when finite-size
effects are present, let us try to develop a simple mathematical model by
considering the differential equation
\begin{equation}
\frac{dy}{d(x^{2})}=-a_{r}y^{r}-(a_{q}-a_{r})y^{q}\;\;\;\;(a_{q}\geq
a_{r}\geq 0;\,q>r;\,y(0)=1)\,.  \label{diffeq2}
\end{equation}%
This equation has very interesting and different solutions depending on the
choice of $r$ and $q$ values (see refs.\cite{tsallis2,bemski,tsatir}),
but for our purpose, let us concentrate on case $r=1$ and $q>1$,
whose solution is given by
\begin{equation}
y=\left[ 1-\frac{a_{q}}{a_{1}}+\frac{a_{q}}{a_{1}}\,e^{(q-1)a_{1}\,x^{2}}%
\right] ^{1/(1-q)}\,.
\label{crossover}
\end{equation}%

If $a_{1}=0$, then the solution coincides with the $q$-Gaussian, whereas if $a_{q}=a_{1}$
(which means that $q=1$), the solution turns out to be the Gaussian. On the other hand,
between these two extremes, namely if $a_{q}>a_{1}>0$ and $q>1$, we obtain a crossover
between them. Specifically, for $(q-1 )\, a_1 \, x^2 \ll 1 $,  Eq.~(\ref{crossover})
approaches a $q$-Gaussian, $y \sim \left[ 1- (1-q ) \, a_q \, x^2 \right] ^{1/(1-q)} $.
Our results, which are depicted in Fig.~3, show that the small values of $a_1$ imply that
the $q$-Gaussian form is valid up to rather large values of $x$. On the other hand,
for $(q-1 ) \, a_1 \, x^2 \gg 1$, the exponential outnumbers the remaining terms leading
to the Gaussian behaviour,
$$y \asymp \left(\frac{a_{q}}{a_{1}}\right) ^{1/(1-q)} \exp \left(- a_1 \, x^2 \right).$$
The approximate dependence of Eq.~(\ref{crossover}) can thus be split into different regions
defined by three values of $x$. Namely the first value is
$$ x_a \sim \sqrt{\frac{W\left[-\frac{a_1}{a_q}\right]}{a_1(1-q)}},$$
($W\left[\ldots \right]$ is the Lambert $W$ function~\cite{functions}) whence the curve
assumes a power-law dependence described by the exponent $2/(q-1)$ that persists up to
$$ x_b \sim \sqrt{ \frac{ \ln 2}{a_1 (q-1) }},$$
when the it starts being perturbed by the Gaussian dependence. Last, there is the final
convergence to the Gaussian functional form which occurs at
$$x_c \sim \sqrt{\frac{\ln \left(1 - \frac{a_1}{a_q}+\frac{a_q}{a_1} \right)}
{a_1 \, (q-1)}}.$$

This crossover seems to coincide with the behavior of the return distributions of
the $N$ dependent cases as plotted with dashed black lines on top of each
curve in Fig.~3(b). This behavior simply reveals that the longer the
power-law regime persists for avalanche size distribution, the better the
appropriate $q$-Gaussian dominates in the return distribution. Finally, as
$N\rightarrow \infty$, the power-law regime prevails for the avalanche size
distribution giving forth a return distribution following the appropriate $q$-Gaussian
for the \textit{entire} region.

%------------------------ Figure 3 --------------------------
\begin{figure*}[t]
\begin{center}
\includegraphics[width=9cm]{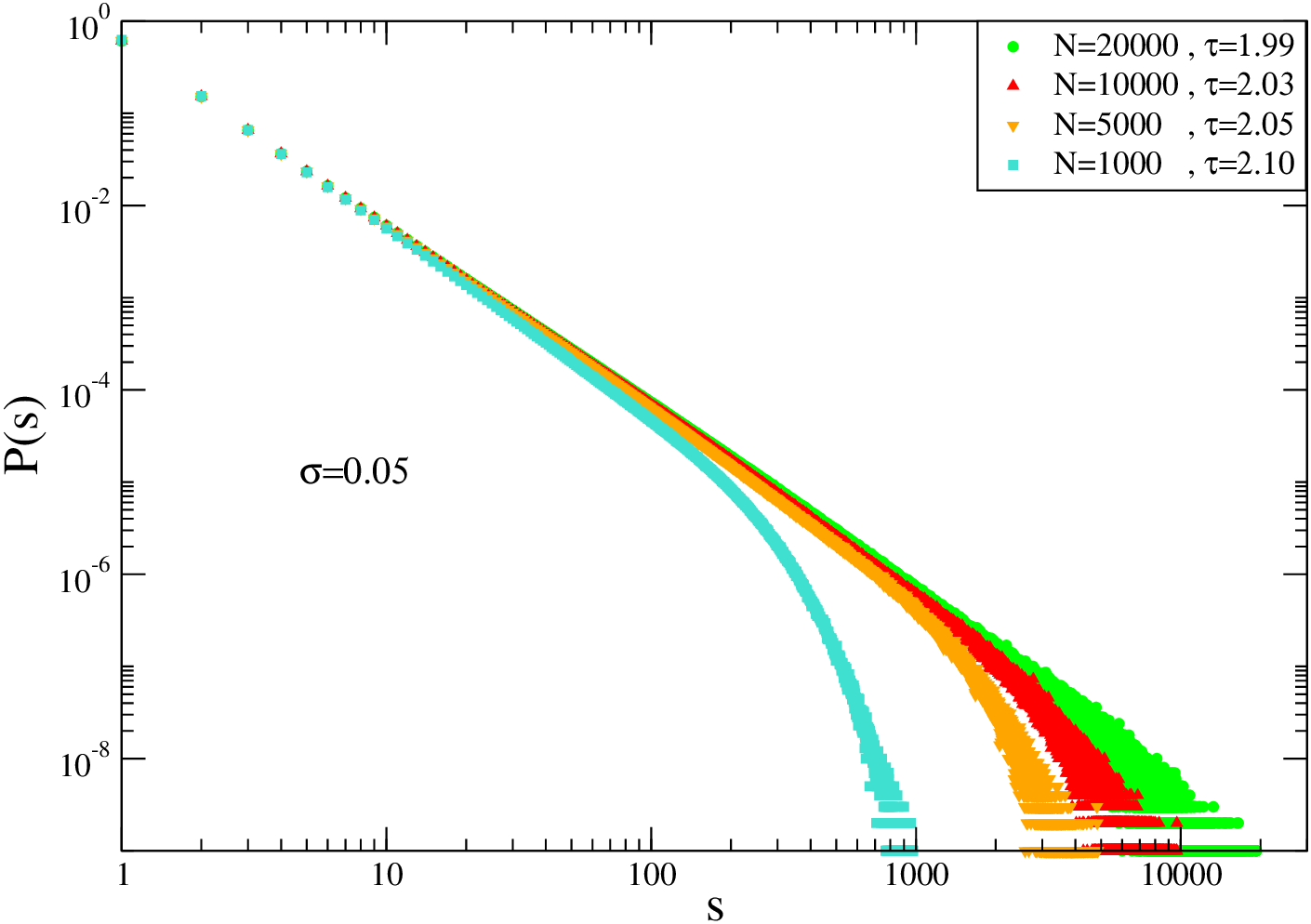} \\[1.2cm]
\includegraphics[width=9cm]{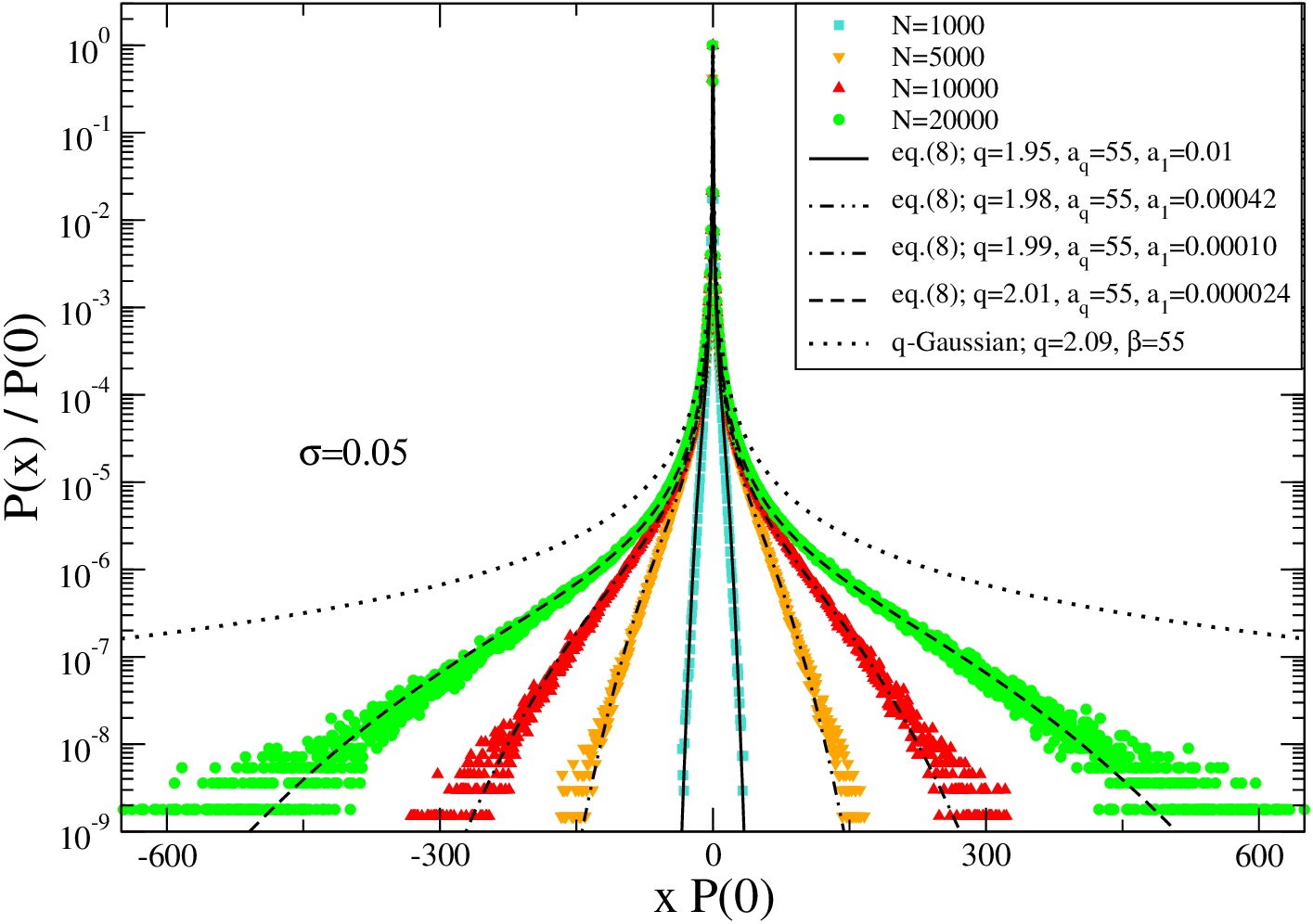}
\end{center}
\caption{(a) Avalanche size distributions for $N$ dependent
case. (b) Corresponding return distributions of the same $N$ values.}
\label{fig:fig3}
\end{figure*}
%-----------------------------------------------------------

%%%%%%%%%%%%%%%%%%%%%%%%%%%%%%%%%%%%%%%%%%

\section{\label{sec:Conc} Conclusion}

%%%%%%%%%%%%%%%%%%%%%%%%%%%%%%%%%%%%%%%%%%
In this work, we have studied the behavior of the return distributions for
the CNM by directly simulating the size independent case.
By means of extensive simulations, it is clearly shown that these
distributions converge to $q$-Gaussians with appropriate $q$ values
which are deduced {\it a priori} from the exact relation (\ref{qvstau2})
that we developed here. It is worth noting that although the $q$-Gaussian
description is actually an analytical approximation the result provides for
an understandable depiction of the distribution, which hardly occurs when
we keep a special functions representation, with no fundamental accuracy lost.
This relation makes the $q$ parameter be related to one of the
well-known exponents (avalanche size exponent $\tau$) of such complex
systems and therefore it rescues $q$ from being a fitting parameter in this
analysis. Moreover, since the model parameter $\sigma$ allows us to obtain
different $\tau$ values, we were able to check this behavior for various
cases. These results clearly imply that the observed behavior is not
restricted to self-organized critical models, but instead it seems to be
a rather generic feature presented by many complex systems which
exhibit asymptotic power-law distribution of avalanche sizes.

We have also investigated the finite-size effect by simulating directly
the model dynamics and found that the convergence to appropriate $q$-Gaussian 
starts from the central part and gradually evolves towards the tails as the 
system size increases. This is in complete agreement with the 
gradual extension of the power-law regime in the avalanche size 
distribution before the appearance of the exponential decay due to finite-size
of the system. These results corroborate the analysis of size independent case
since it is clearly seen that, as $N\rightarrow\infty$, curves of return
distributions for size dependent case converge to the one comes from the
size independent case.

Finally it should be noted that, since it is generically extremely difficult
(if not impossible) to achieve the size independent case for such complex
systems, the size dependent case has its particular importance.
Therefore, although the return distributions appear to be $q$-Gaussians
for the entire region in the thermodynamic limit, we have tried to propose
a mathematical model in order to explain the behavior of return distributions
for the size dependent case.

%%%%%%%%%%%%%%%%%%%%%%%%%%
\section*{Acknowlegment}
%%%%%%%%%%%%%%%%%%%%%%%%%%

We are indebted to M E J Newman for providing us his fast (size independent)
code for the coherent noise model and C Anteneodo for interesting discussions
about passage problems and related references.
This work has been supported by TUBITAK (Turkish Agency) under the Research Project
number 104T148 and by Ege University under the Research Project number 2009FEN027.

\appendix

\section{The CNM master equation}

The dynamics of the CNM can be described according to the probability of
having $n$ agents in the system that at time $t$ present a critical value up
to $x$, $P_{n}\left( x,t\right) $. In conformity with step 1 we can write
the master-equation,
\begin{eqnarray}
P_{n}\left( x,t\right)  &=&P_{n}\left( x,t-1\right) W_{n\rightarrow
n}+\sum\limits_{l=1}^{n}P_{n-l}\left( x,t-1\right) W_{n-l\rightarrow
n}\nonumber \\
&&+\sum\limits_{l=1}^{N-n}P_{n+l}\left( x,t-1\right) W_{n+l\rightarrow n}-
\sum\limits_{l=1}^{n}P_{n}\left( x,t-1\right) W_{n\rightarrow
n-l}\nonumber \\
&&-\sum\limits_{l=1}^{N-n}P_{n}\left( x,t-1\right) W_{n\rightarrow n+l}
\end{eqnarray}
with the probability transitions $W$ given by
\begin{equation}
W_{n\rightarrow n}=x^{n}\,F^{\prime }\left( x\right)
+\sum_{m=1}^{n}\int_{0}^{x}x^{m}\frac{m}{n}\,p_{thresh}\left( \eta \right)
\,\delta \left( m-M_{x,\eta }\right) d\eta ,
\end{equation}
where the first term on the rhs comes from the case $\eta \geq x$ and the
second one otherwise.
The inverse cumulative probability $F^{\prime }\left(
x\right) \equiv 1-F\left( x\right) =1-\int_{0}^{x}p_{thresh}\left( z\right)
\,dz$~\footnote{For our case, i.e., $p_{thresh}(z)= \sigma ^{-1} \exp[- z/ \sigma]$
implies $F(z) = 1- \exp[- z / \sigma]$.}
and $M_{x,\eta }\equiv \sum_{i=1}^{N}\Theta \left[ \eta -x_{i}\left(
t\right) \right] \Theta \left[ x-\eta \right] $ means the number of agents
with critical value below $x$ and $\eta$. The following elements are

\begin{equation}
W_{n-l\rightarrow n}=W_{n\rightarrow n+l}=0 ,
\end{equation}

\begin{eqnarray}
%\fl 
W_{n+l\rightarrow n} =\left(
\begin{array}{c}
n+l \\
n%
\end{array}%
\right) x^{n}\left( 1-x\right) ^{l}\,F^{\prime }\left( x\right) + \nonumber \\
\sum_{m\geq l}^{n+l}\int_{0}^{x}\left(
\begin{array}{c}
m \\
m-l%
\end{array}%
\right) x^{m-l}\left( 1-x\right) ^{l}\frac{m}{n+l}\,p_{thresh}\left( \eta
\right) \,\delta \left( m-M_{x,\eta }\right) d\eta ,
\end{eqnarray}

\begin{eqnarray}
%\fl 
W_{n\rightarrow n-l} =\left(
\begin{array}{c}
n \\
n-l%
\end{array}%
\right) x^{n-l}\left( 1-x\right) ^{l}\,F^{\prime }\left( x\right) + \nonumber \\
\sum_{m\geq l}^{n}\int_{0}^{x}\left(
\begin{array}{c}
m \\
m-l%
\end{array}%
\right) x^{m-l}\left( 1-x\right) ^{l}\frac{m}{n}\,p_{thresh}\left( \eta \right)
\,\delta \left( m-M_{x,\eta }\right) d\eta .
\end{eqnarray}

This corresponds to a matrix with vanishing elements below the diagonal.
From these relations is then possible to spell out the occurrence of an
avalanche of size $s$
\begin{equation}
P\left( s\right) =\left\{
\begin{array}{ccc}
\sum_{n=0}^{N}\int_{0}^{1}\frac{dP_{n}\left( x,t-1\right) }{dx}%
\,W_{n\rightarrow n}\,dx & \Leftarrow  & s=0 \\
&  &  \\
\begin{array}{c}
\sum_{n=0}^{N-s}\int_{0}^{1}\frac{dP_{n+s}\left( x,t-1\right) }{dx}%
\,W_{n+s\rightarrow n}\,dx+ \\
\sum_{n=s}^{N}\int_{0}^{1}\frac{dP_{n}\left( x,t-1\right) }{dx}%
\,W_{n\rightarrow n-s}\,dx%
\end{array}
& \Leftarrow  & s\neq 0%
\end{array}%
\right. ,
\end{equation}
which is numerically well described by the power-law~(\ref{ps}) with a small
value of $\varepsilon $.

Regarding step 2 the master equation is abstractly pretty much the same,
\begin{eqnarray}
%\fl 
P_{n}\left( x,t+1\right) =& P_{n}\left( x,t\right) W_{n\rightarrow
n}+\sum\limits_{l=1}^{n}P_{n-l}\left( x,t\right) W_{n-l\rightarrow
n}+\sum\limits_{l=1}^{N-n}P_{n+l}\left( x,t\right) W_{n+l\rightarrow n}
\nonumber \\
&-\sum\limits_{l=1}^{n}P_{n}\left( x,t\right) W_{n\rightarrow
n-l}-\sum\limits_{l=1}^{N-n}P_{n}\left( x,t\right) W_{n\rightarrow n+l}\,,
\end{eqnarray}
with the probability transition matrix is given by
\begin{eqnarray}
%\fl 
W_{n\rightarrow n} =\int_{0}^{1}\left(
\begin{array}{c}
n \\
\rho f\,N%
\end{array}%
\right) \left(
\begin{array}{c}
N-n \\
\left( 1-\rho \right) f\,N%
\end{array}%
\right) \left(
\begin{array}{c}
f\,N \\
\rho f\,N%
\end{array}%
\right) \left( x\frac{n}{N}\right) ^{\rho \,f\,N} \,  \times \nonumber \\
\left[ \left( 1-x\right) \left( 1-\frac{n}{N}\right) \right]
^{\left( 1-\rho \right) fN}
 \Theta \left[ n-\rho fN\right]\;
\Theta \left[N-n-\left( 1-\rho \right) fN\right] \,d\rho
\end{eqnarray}
where $\rho $ is used to define the subfraction of agents, $\rho f N$,
whose critical value before updating was less than $x$.

\begin{eqnarray}
%\fl 
W_{n-l\rightarrow n} =\int_{0}^{1}\left(
\begin{array}{c}
n-l \\
\rho f\,N%
\end{array}%
\right) \left(
\begin{array}{c}
N-\left( n-l\right)  \\
\left( 1-\rho \right) f\,N%
\end{array}%
\right) \left(
\begin{array}{c}
f\,N \\
\rho f\,N+l%
\end{array}%
\right) \times  \nonumber \\
 \left( \frac{n-l}{N}\right) ^{\rho \,f\,N}\left( 1-\frac{n-l}{N}%
\right) ^{\left( 1-\rho \right) fN}x^{\rho \,f\,N+l}\left( 1-x\right)
^{\left( 1-\rho \right) fN-l}\times  \nonumber \\
 \Theta \left[ n-l-\rho fN\right] \,\,\Theta \left[ N-\left(
n-l\right) -\left( 1-\rho \right) fN\right] \,\,d\rho
\end{eqnarray}

\begin{eqnarray}
%\fl 
W_{n+l\rightarrow n} =\int_{0}^{1}\left(
\begin{array}{c}
n+l \\
\rho f\,N%
\end{array}%
\right) \left(
\begin{array}{c}
N-\left( n+l\right)  \\
\left( 1-\rho \right) f\,N%
\end{array}%
\right) \left(
\begin{array}{c}
f\,N \\
\rho f\,N-l%
\end{array}%
\right) \times  \nonumber \\
 \left( \frac{n+l}{N}\right) ^{\rho \,f\,N}\left( 1-\frac{n+l}{N}%
\right) ^{\left( 1-\rho \right) fN}x^{\rho \,f\,N-l}\left( 1-x\right)
^{\left( 1-\rho \right) fN+l}\times  \nonumber \\
 \Theta \left[ n+l-\rho fN\right] \,\,\Theta \left[ N-\left(
n+l\right) -\left( 1-\rho \right) fN\right] \,\,d\rho
\end{eqnarray}

\begin{eqnarray}
%\fl 
W_{n\rightarrow n-l} =\int_{0}^{1}\left(
\begin{array}{c}
n \\
\rho f\,N%
\end{array}%
\right) \left(
\begin{array}{c}
N-n \\
\left( 1-\rho \right) f\,N%
\end{array}%
\right) \left(
\begin{array}{c}
f\,N \\
\rho f\,N-l%
\end{array}%
\right) \times  \nonumber \\
 \left( \frac{n}{N}\right) ^{\rho \,f\,N}\left( 1-\frac{n}{N}\right)
^{\left( 1-\rho \right) fN}x^{\rho \,f\,N-l}\left( 1-x\right) ^{\left(
1-\rho \right) fN+l}\,\times  \nonumber \\
 \Theta \left[ n-\rho fN\right] \,\,\Theta \left[ N-n-\left( 1-\rho
\right) fN\right] \,d\rho ;
\end{eqnarray}

\begin{eqnarray}
%\fl 
W_{n\rightarrow n+l} =\int_{0}^{1}\left(
\begin{array}{c}
n \\
\rho f\,N%
\end{array}%
\right) \left(
\begin{array}{c}
N-n \\
\left( 1-\rho \right) f\,N%
\end{array}%
\right) \left(
\begin{array}{c}
f\,N \\
\rho f\,N+l%
\end{array}%
\right)  \times \nonumber \\
\left( \frac{n}{N}\right) ^{\rho \,f\,N}\left( 1-\frac{n}{N}\right)
^{\left( 1-\rho \right) fN}x^{\rho \,f\,N+l}\left( 1-x\right) ^{fN\left(
1-\rho \right) -l} \times \nonumber \\
\Theta \left[ n-\rho fN\right] \,\,\Theta \left[ N-n-\left( 1-\rho \right)
fN\right] \,d\rho .
\end{eqnarray}

%%%%%%%%%%%%%%%%%%%%%%%%%%%%%%

\end{document}